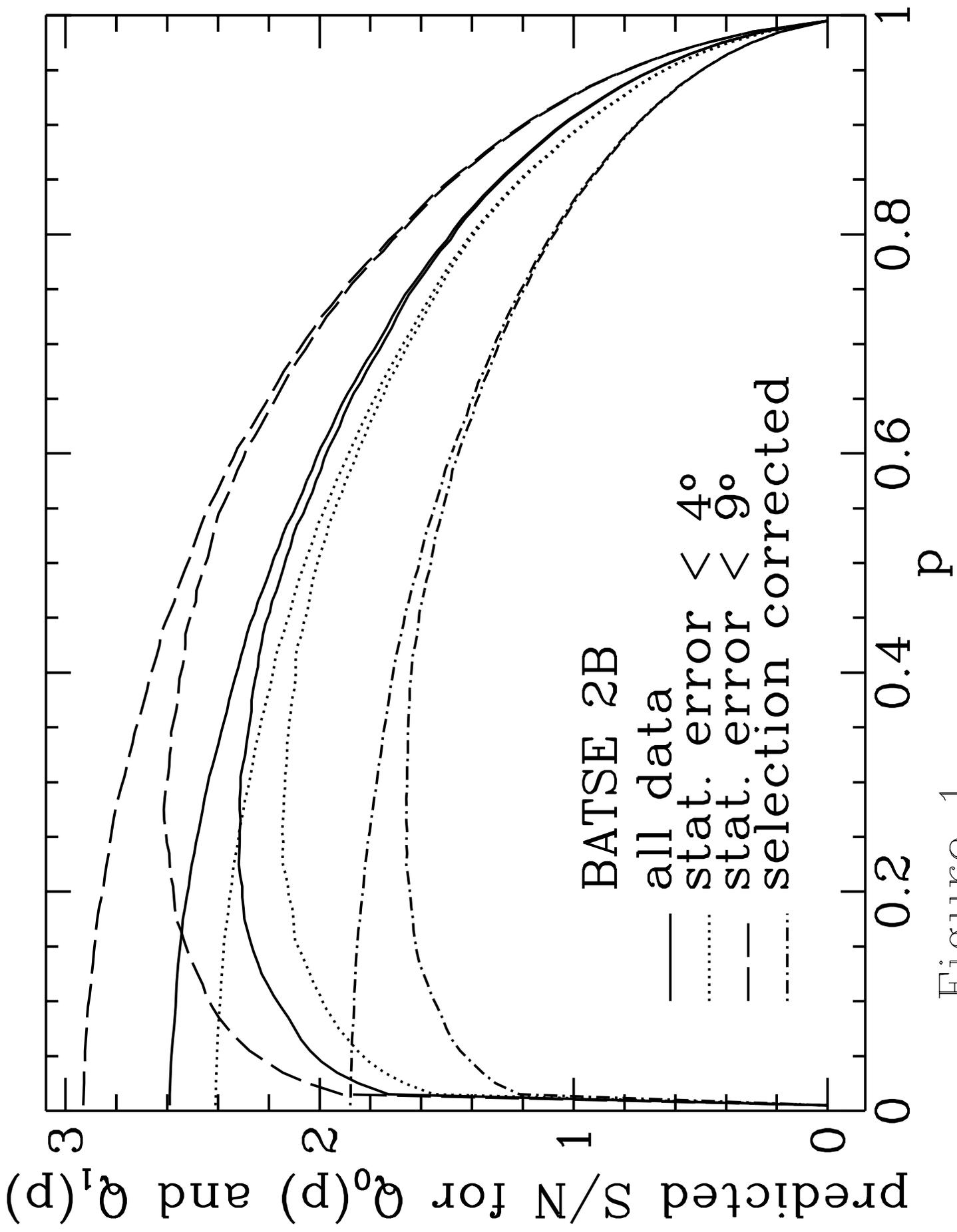

Figure 1

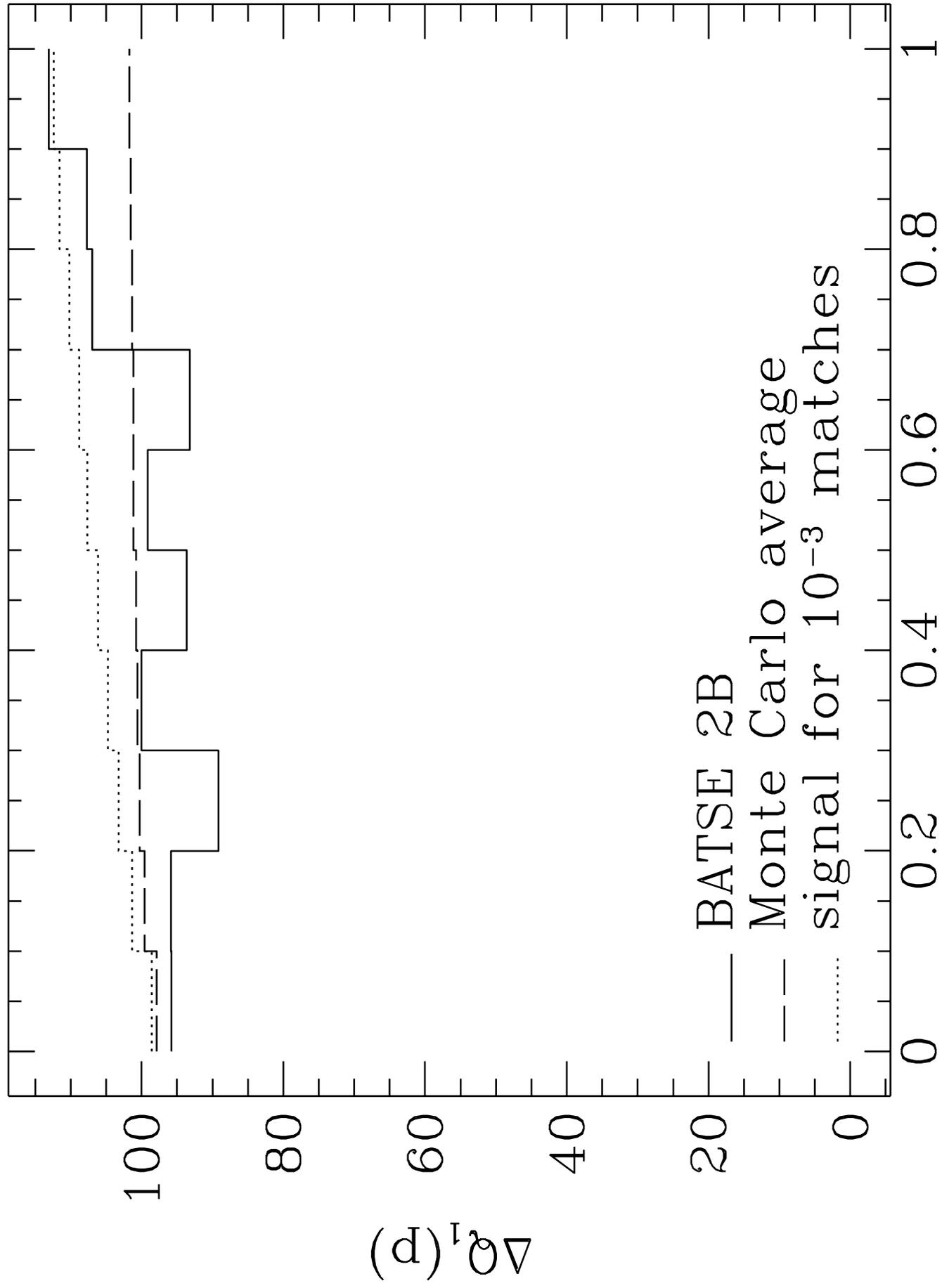

Figure 2(a)

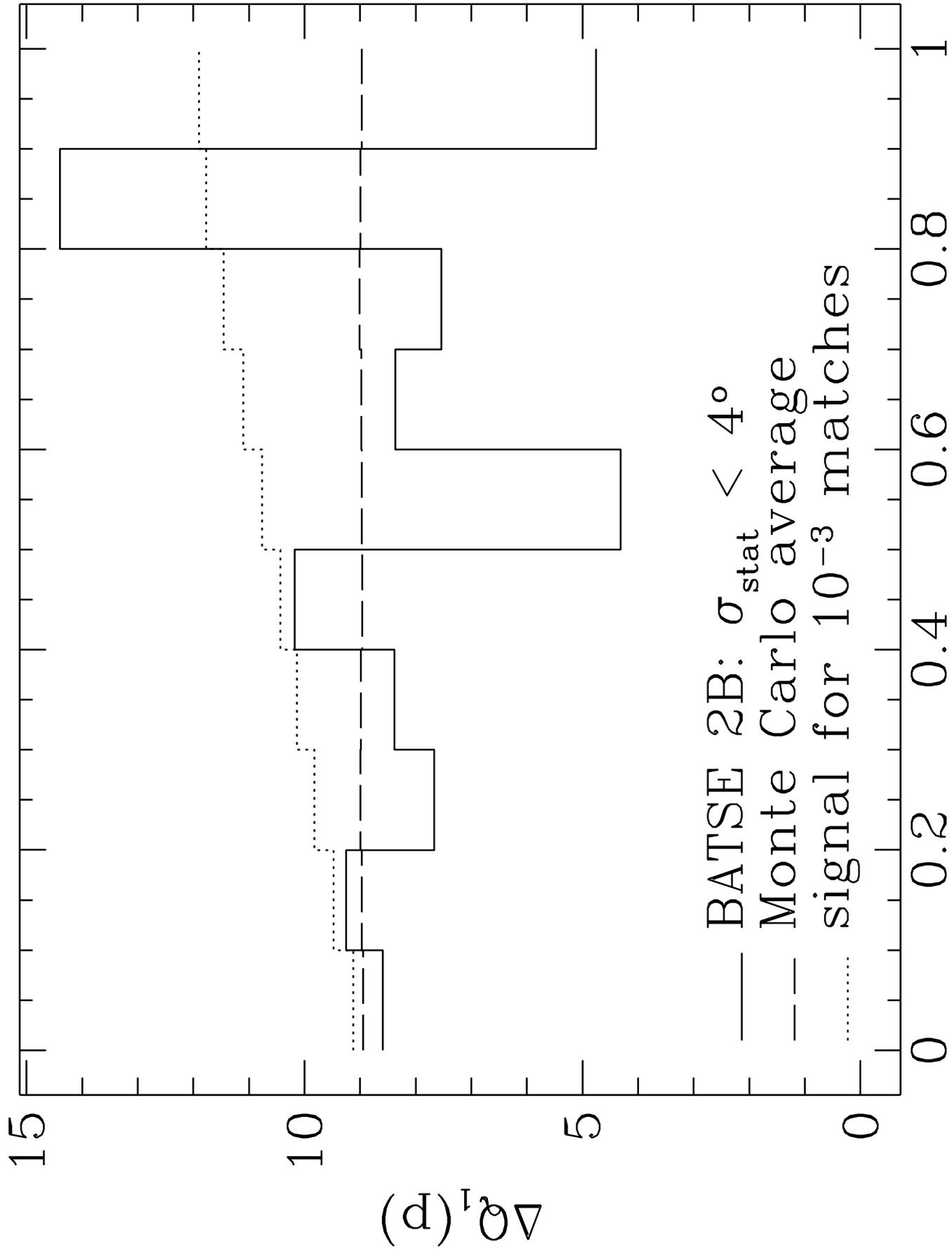

Figure 2(b)

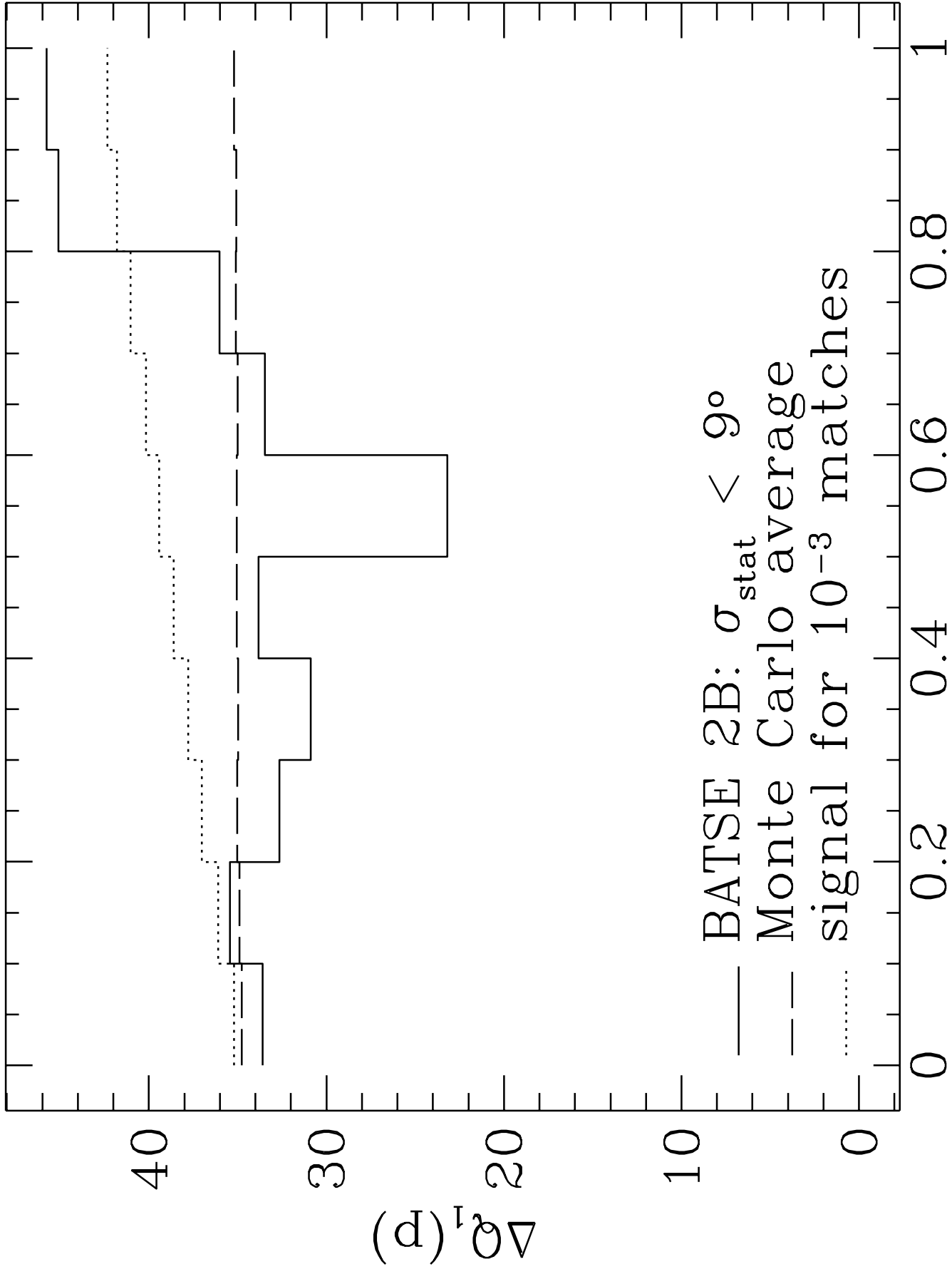

Figure 2(c)

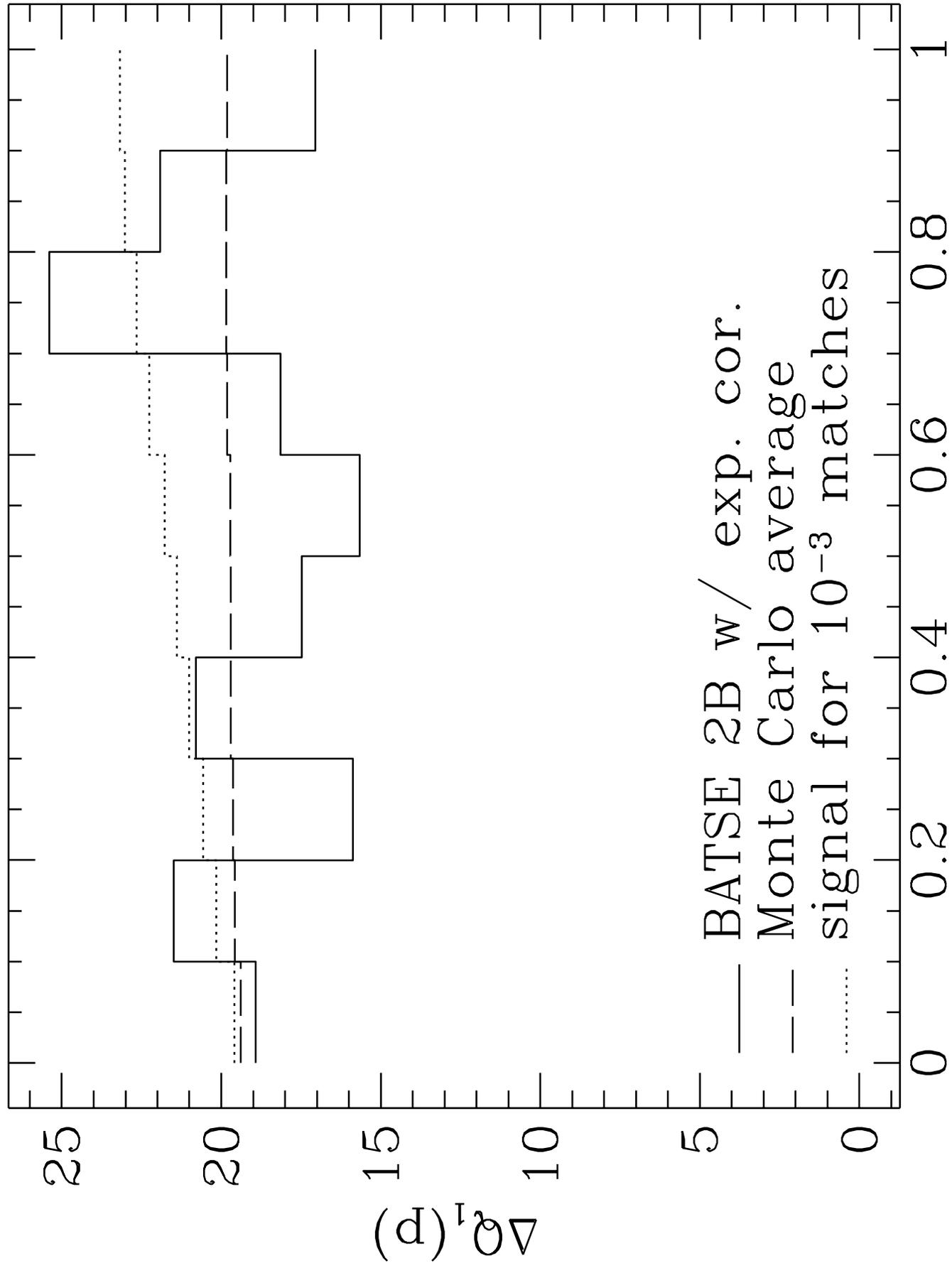

Figure 2(d)

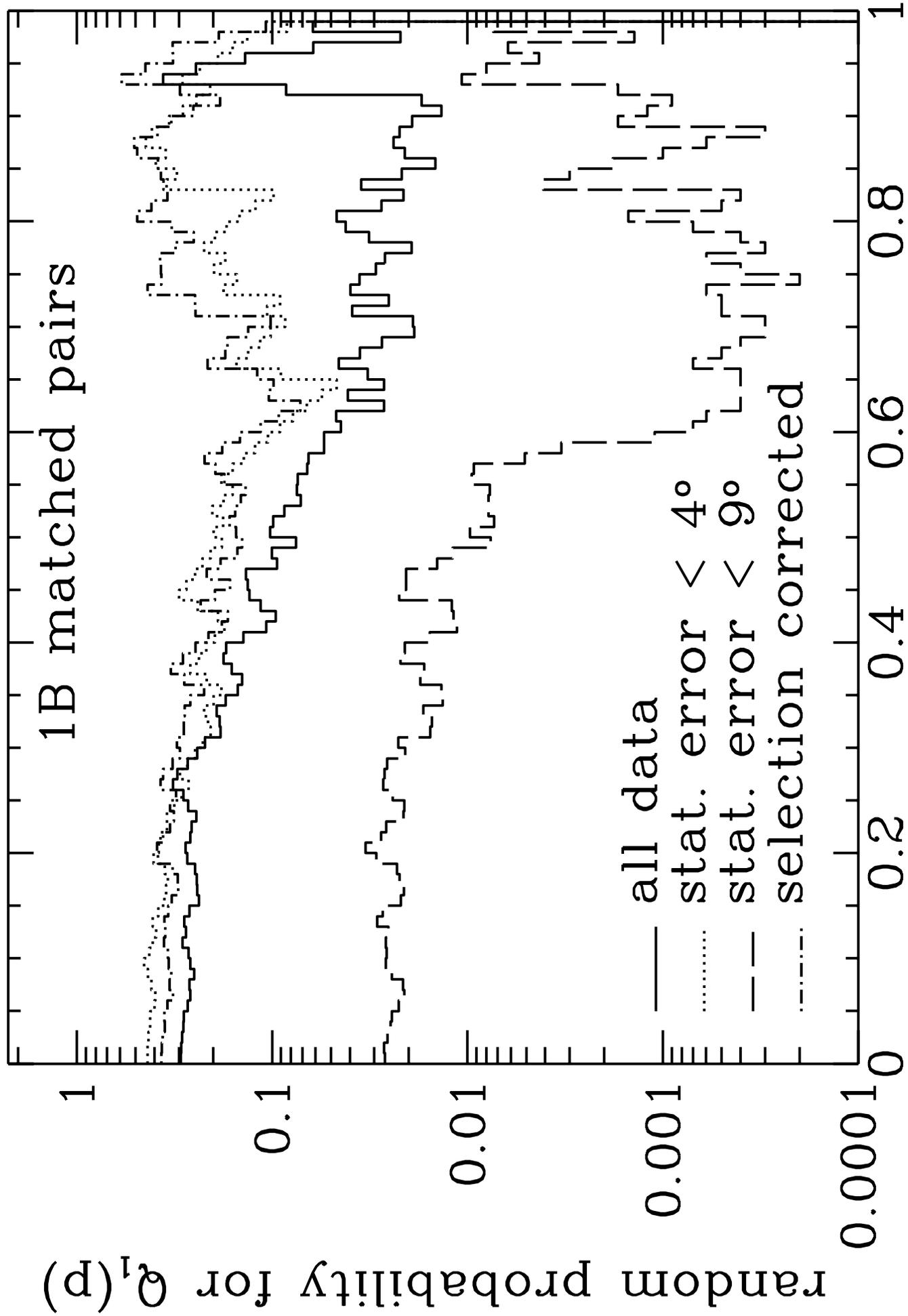

Figure 3(a)

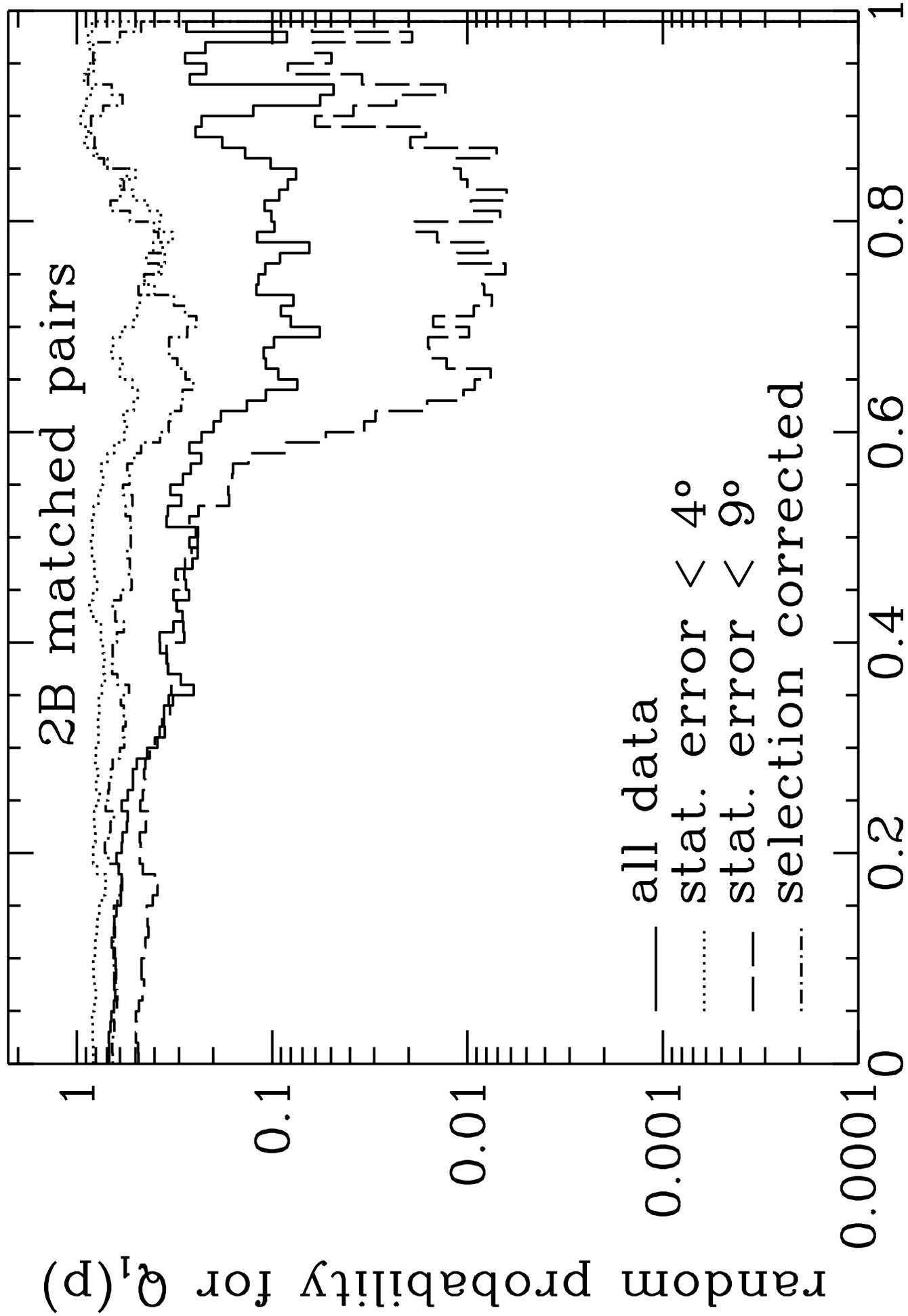

Figure 3(b)

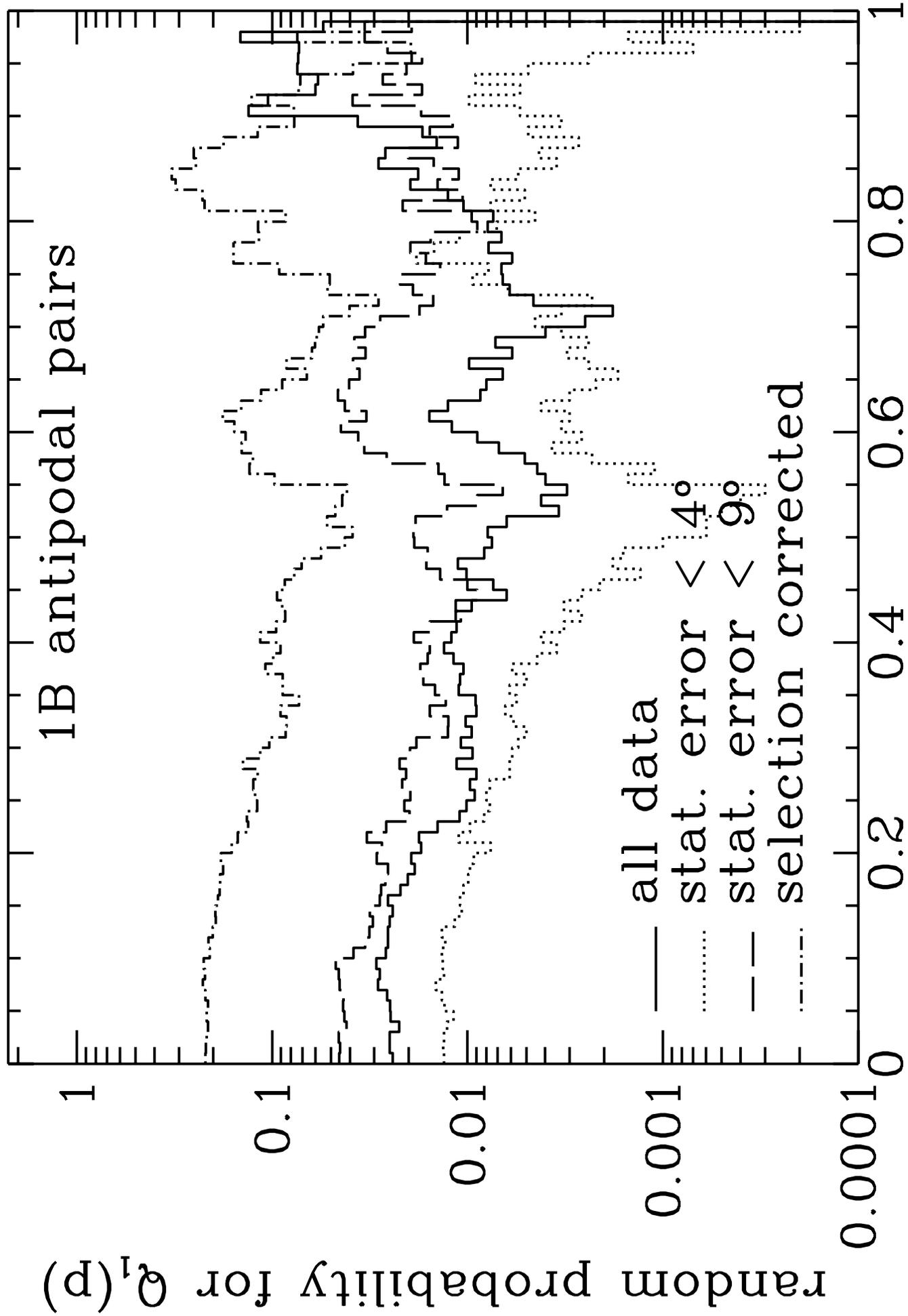
Figure 3(c)

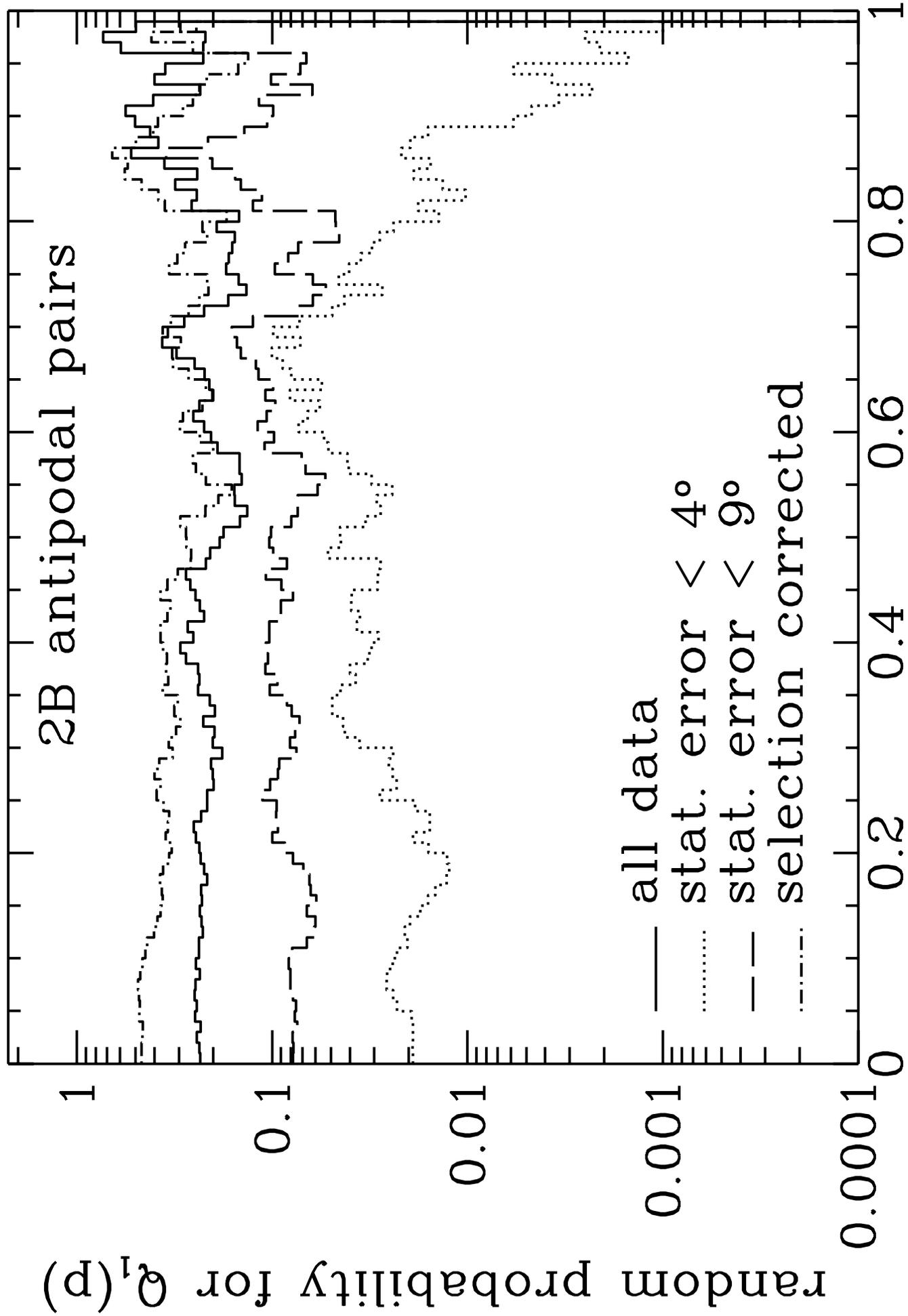

Figure 3(d)

# Is There Evidence for Repeating Gamma Ray Bursters in the BATSE Data?

DAVID P. BENNETT AND SUN HONG RHIE

*Center for Particle Astrophysics, University
of California, Berkeley, CA 94720*

and

*Institute of Geophysics and Planetary Physics, Lawrence
Livermore National Laboratory, Livermore, CA 94550*

## ABSTRACT

The possibility that classical gamma ray bursts (GRB) occasionally repeat from the same locations on the sky provides a critical test of GRB models. There is currently some controversy about whether there is evidence for burst repetition in the BATSE data. We introduce a gamma ray burst "pair matching" statistic that can be used to search for a repeater signal in the BATSE data. The pair matching statistic is built upon the reported position errors for each burst and is more sensitive than previously used statistics at detecting faint repeating bursts or multiple burst recurrences. It is also less likely to produce (false) evidence of burst repetition due to correlations in the positions that are inconsistent with repeating bursters. We find that the excesses in "matched" and "antipodal" pairs seen with other statistics are caused by an excess of pairs with a separation smaller than their error bars would indicate. When we consider all separations consistent with the error bars, no significant signal remains. We conclude that the publicly available BATSE 1B and 2B data sets contain no evidence for repeating gamma ray bursters.

# 1. Introduction

Since gamma ray bursts were first discovered in 1967 (Klebesadel, Strong, and Olson, 1973), their origin has remained a mystery. Many models have been proposed to explain them, but just prior to the launch of the Compton Gamma Ray Observatory (CGRO) there seemed to be a general (though not complete) consensus that galactic neutron stars must be the source of the bursts (Liang, and Petrosian, 1986, Hurley, 1989, and Harding, 1991). This consensus was shaken by the first data from the BATSE detector (Meegan, *et. al.*, 1992) which saw an isotropic distribution of gamma ray bursts with a dearth of faint bursts. The most straight forward interpretation of these results is that BATSE is seeing to the edge of an isotropic burst distribution contrary to the predictions of the galactic models. Attempts were made to try to rescue the galactic neutron star models. These have generally had some difficulty with explaining the BATSE data (Hartmann, 1994, and references therein), but a recent paper by Podsiadlowski, Rees, and Ruderman (1994) suggests that some extended galactic halo models may still fit the BATSE data. Nevertheless, the BATSE results have shifted the consensus view toward models in which the bursters are at cosmological distances.

One other piece of evidence that has some bearing on the question as to whether gamma ray bursts are galactic or cosmological is the question of whether the bursting is a recurrent activity of the burst sources. This might be expected in galactic halo models where a typical gamma ray burst emits $\sim 10^{43}$ergs, but is difficult to reconcile with the cosmological models which generally require so much energy ($\sim 10^{51}$ergs) that the burst source is typically consumed in the process of generating a burst. If GRB's were found to repeat from the same sources, then this would be evidence that they have a galactic origin after all.

The first claim of statistical evidence for repeating bursters in the BATSE data was made by Quashnock and Lamb (1993, QL hereafter). QL analyzed the BATSE 1B burst catalog of 260 events and found an excess of bursts with a "nearest neighbor" burst within $\lesssim 4°$ compared to expectation for a random distribution of burst locations. One puzzling feature of their result was that the clustering occurs on a scale somewhat smaller than the typical burst position error estimates. QL interpreted these results to imply that some of the bursts were due to multiple repetitions from the same sources so that the nearest neighbor would be closer together than "typical" bursts from the same source. QL also identified the bursts responsible for their signal to be mostly short and faint events (QL's 'type II' events).

In response to QL, a number of authors have analyzed the angular distribution of bursts using the two point angular correlation function of the bursts. Narayan and Piran (1993, NP hereafter) have found an apparently statistically significant correlation at small angular scales in agreement with QL's result, but they also find a signal of similar significance at angular separations $\gtrsim 176°$. They also find that the large angle correlations can be seen with a "farthest neighbor" statistic analogous to Quashnock and Lamb's "nearest neighbor" statistic. Angular correlation function analyses have also been carried out by Blumenthal, *et. al.*(1993), and Hartmann, *et. al.*(1993), who find that apparent signals in the angular correlation function seem to go away when larger samples of gamma ray bursts are analysed.

Here, we present a new statistic designed to detect statistical evidence of gamma ray burst recurrence in the BATSE catalog. This statistic is motivated by the simple observation that the separations of bursts which originate from the same source should be determined solely by the position measurement errors. In other words, the position differences of bursts from the same source constitute an *error distribution*. Therefore, the set of burst positions from the BATSE catalog can be considered as a mixture of *error distributions* and a background *isotropic distribution*. The error distribution is highly localized about zero separation while the isotropic distribution is not. We use this difference to build a burst pair "match probability" statistic, and we then show that this statistic is generally more sensitive to weak burst repetition signals than the correlation function or nearest neighbor statistics used by other authors. Finally, we apply this statistic to the BATSE 1B and 2B data sets and show that distribution of BATSE burst positions does not contain a significant indication of burst repetition.

## 2. The Pair Matching Statistic

One serious drawback of both the "nearest neighbor" statistic and the angular correlation function is that they ignore the individual angular position error estimates that are provided with the BATSE data. Thus, a pair of bursts separated by 20° will give the same signal if the position errors are each 4° or 20° even though the chance that the bursts come from the same source should be quite different in these two cases. If we want to test whether bursts are recurring from the same positions on the sky, we should ask for each pair of bursts how likely it is for the pair to have originated from the same position on the sky. Therefore, let's define the "match probability" for bursts $i$ and $j$. If $\theta_{ij}$ is the angular distance between bursts $i$ and $j$, the joint probability density of the pair is

$$\frac{1}{2\pi\sigma_{ij}^2} e^{-\theta_{ij}^2/2\sigma_{ij}^2}$$

where $\sigma_{ij}^2 = \sigma_{i,\mathrm{stat}}^2 + \sigma_{i,\mathrm{sys}}^2 + \sigma_{j,\mathrm{stat}}^2 + \sigma_{j,\mathrm{sys}}^2$ is the quadrature sum of the statistical and systematic position errors for bursts $i$ and $j$. (Note that the errors given in the BATSE catalog are two dimensional 68% confidence level radii which are larger than the Gaussian $\sigma$ values by a factor of 1.5096.) We now define the "match probability" for bursts $i$ and $j$ to be

$$p_{ij} = \frac{1}{\sigma_{ij}^2} \int_{\theta_{ij}}^{\infty} e^{-\theta^2/2\sigma_{ij}^2}\, \theta\, d\theta = e^{-\theta_{ij}^2/2\sigma_{ij}^2}\ . \tag{2.1}$$

which is just the formal probability that if bursts $i$ and $j$ originated from the same position on the sky, they would have a separation at least as large as that observed. Thus, for burst pairs that come from the same source, we should expect $p_{ij}$ to be uniformly distributed between 0 and 1. The formula (2.1) implicitly assumes that the burst separations and error estimates are small so that the integral can be done on a flat surface rather than a sphere. This will lead us to overestimate the probability of large separations for bursts with relatively large error bars. However, the error distribution for the bursts with the largest errors is known to be non-Gaussian (Fishman, *et. al.*1993), so this is unlikely to make things appreciably worse.

One can express the expected distribution of $p_{ij}$ values as a distribution function: $g(p)$. For matched pairs, $g(p)$ is just a constant, so for $N_{\mathrm{matched}}$ matched pairs, we have

$$g_{\mathrm{matched}}(p) = N_{\mathrm{matched}}\ , \tag{2.2}$$

while for a set of $N_{\mathrm{random}}$ randomly distributed burst pairs we have

$$g_{\mathrm{random}}(p) = \sum_{i,j<i} \frac{\sigma_{ij}^2}{2p} = \frac{N_{\mathrm{random}} \left\langle \sigma_{ij}^2 \right\rangle}{2p}\ , \tag{2.3}$$

where we have used the small angle approximation to derive eq. (2.3). Note that $g_{\mathrm{random}}(p)$ does not actually diverge for small $p$ because there is a minimum allowed value of $p$ which corresponds to $\theta = \pi$. In order to test for repeating bursts, we will need a statistic that will be able to detect a signal of the form, (2.2), with a background of 'noise' of the form (2.3), with $N_{\mathrm{random}} \gg N_{\mathrm{matched}}$.

We can use eq. (2.1) to define such a set of statistics which measure the number of position matches in the BATSE catalog:

$$Q_\alpha(p) = \sum_{\text{pairs}} p_{ij}^\alpha \Theta(p_{ij} - p) , \qquad (2.4)$$

where $\Theta$ is the familiar step function. The value of $Q_\alpha(p)$ for a completely random and uncorrelated data set will depend on the number of data points and the distribution of error bars. Therefore, we will need to do Monte Carlo simulations of data sets with the observed distribution of error bars and random positions in order to assess the statistical significance of these statistics. If there is a population of gamma ray bursters that does repeat, then the expected signal is given by

$$Q_\alpha(p) - \langle Q_\alpha(p) \rangle_{\text{RANDOM}} = \frac{N_{\text{matched}}}{\alpha + 1} \left(1 - p^{\alpha+1}\right) - \frac{(1 - p^\alpha)}{2\alpha} \sum_{\substack{\text{matched} \\ \text{pairs}}} \sigma_{ij}^2 , \qquad (2.5)$$

where $\langle Q_\alpha(p) \rangle_{\text{RANDOM}}$ is the value determined by Monte Carlo simulations of BATSE burst catalogs with positions that have been reassigned random values, and $N_{\text{matched}}$ is the number of burst pairs originating from the same physical burst location on the sky. For $\alpha = 0$, $(1 - p^\alpha)/\alpha$ is replaced by $-\ln p$ in the last term of eq. (2.5). Using eq. (2.5) we can determine the the signal to noise expected for the statistic $Q_\alpha(p)$ as a function of $p$ by comparing the expected signal to the RMS fluctuation about $\langle Q_\alpha(p) \rangle_{\text{RANDOM}}$. This is shown in Figure 1 for the full BATSE 2B data set and a number of subsets that we will discuss below. The curves are normalized to the expected signal if 1 pair of bursts out of 1000 originated from the same position on the sky. If the repetition time for any repeating bursts is small enough so that multiple bursts are seen in a given sample, then the fraction of bursts pairs that originated from the same source should remain constant while the uncertainty in this number scales as the square root of the total number of pairs. On the other hand, if the timescale of burst repetition is such that some of the repeating bursters are seen only once in a data set, then we would expect that the fraction of repeating pairs would increase as the data set is extended in time. In this later case, the pair matching signal-to-noise should increase faster than the square root of the number of burst pairs. Thus, since the BATSE 2B data set has almost twice as many bursts (with position error estimates) as the BATSE 1B data set, we should expect that a pair matching signal due to repeating bursts should have twice the signal-to-noise in the 2B data set as in the 1B data set.

We should emphasize that although the value of $p$ in eqns. (2.1) and (2.4) is a calculated assuming Gaussian errors, the pair matching statistic *does not* rely upon the assumption that the errors are Gaussian distributed. In this regard it is important that we are testing against the "null hypothesis" of uncorrelated bursts so that we do not have to put realistic correlated burst positions into the model. To the extent that the position errors do not follow a Gaussian distribution, we can consider $p_{ij}$ to be a parameter which is *correlated with* rather than equal to the true match probability. This is particularly obvious for $Q_0(p)$ since it gives no weight to the $p_{ij}$ values inside the sum. In this case, we would just consider $p$ to be a parameter which is well correlated with the true, non-Gaussian match probability. For the other values of $\alpha$, we would just be weighting the sum by a parameter which correlates strongly with the true match probability. The one thing that is affected by the Gaussian assumption is the signal to noise estimates shown in Fig. 1. If the errors are Gaussian, then Fig. 1 indicates that for each data set, $Q_1(0)$ yields the most sensitive measure of the number of matched pairs of bursts. However, since it is known that the errors do have non-Gaussian tails, one might worry that $Q_1(0)$ gives too much weight to small values of $p_{ij}$ where the deviation from Gaussianity is large. This would not mean that $Q_1(0)$ would give biased results, but only that they would be noisier than expected. (An anti-correlation between $p_{ij}$ and the true match probability would be required to bias the results.) Because of the weighting by $p_{ij}$ in the sum (2.4), the small $p_{ij}$ values do not make much contribution to $Q_1(0)$, so this is not an important effect.

Another point worth emphasizing is that it is important to specify the value(s) of $p$ to be used for comparison with data in a way that does not depend on how $Q_1(p)$ depends on $p$ for the actual data used for the comparison. Specifying $p$ "after the fact" is a classic example of *a posteriori* statistics, and it substantially increases the chances of spuriously rejecting the null hypothesis of no matches. *A posteriori* biases are notoriously difficult to correct for (in general) and should be avoided when possible. Therefore, we will base our statistical inferences on the $Q_1(0)$ statistic which we have selected on the *a priori* signal to noise criteria discussed above. The only exception will occur in the next section where we will use $Q_0(p \approx 0.3)$ because noise in this statistic can be calculated more simply.

# 3. Comparison with Other Statistics

Let us now compare the pair matching statistic to the nearest neighbor statistic and correlation functions that have been used by previous authors (QL, NP, Blumenthal, *et. al.*, 1993, Hartmann, *et. al.*, 1993). In order to make this comparison, we have constructed artificial data sets with 260 and 485 artificial bursts. (These are the numbers of bursts with position error estimates in the BATSE 1B and 2B data sets.) Each 260 burst artificial data set has 60 burst pairs originating from the same source locations while each 485 burst data set has 120 such burst pairs. For simplicity, we have assumed that the rms position errors are 8.06° which corresponds to BATSE statistical position error of 7° with at 4° systematic error added in quadrature. This value is chosen because it seems reasonably close to the typical error estimate for the bursts which give rise to the apparent signals seen by the nearest neighbor and correlation function statistics in the BATSE 1B dataset.

| # of bursts | matched pairs | $\theta_c$ | $p_c$ | S/N | $P_{\rm RANDOM}$ |
|---|---|---|---|---|---|
| 260 | 60 | 4° | 0.869 | 1.23 | 0.109 |
| 260 | 60 | 8° | 0.571 | 2.01 | 0.0222 |
| 260 | 60 | 12° | 0.283 | 2.25 | 0.0122 |
| 260 | 60 | 16° | 0.106 | 2.10 | 0.0179 |
| 485 | 120 | 4° | 0.869 | 1.31 | 0.095 |
| 485 | 120 | 8° | 0.571 | 2.15 | 0.0158 |
| 485 | 120 | 12° | 0.283 | 2.40 | 0.0082 |
| 485 | 120 | 16° | 0.106 | 2.25 | 0.0122 |

*Table 1.* The signal to noise ratio expected in the artificial data sets for given numbers of matched pairs using the correlation function and $Q_0(p_c)$ statistics. All the members of the "matched pairs" are assumed to have rms position errors of 8.06° corresponding to a BATSE statistical error of 7°.

Since our matched bursts all have identical position errors, there is a single relationship between $\theta_{ij}$ and $p_{ij}$ for all matched burst pairs. This means that the correlation function value averaged over all separations $< \theta_c$ is identical (up to a normalization factor) to $Q_0(p_c)$ where $p_c = e^{-\theta_c^2/2\sigma^2}$. For a sample of $N$ bursts, the number of burst pairs which fall within $\theta_c$ of each other is $\frac{1}{4}N(N-1)(1-\cos\theta_c)$ (Scott and Tout, 1989). This gives us the information we need to evaluate the signal to noise for the correlation function and $Q_0(p_c)$ for these artificial data sets. The results are shown in Table 1. $P_{\text{RANDOM}}$ is the probability of obtaining a (positive) signal at least as significant as the expected signal from a dataset with completely random burst positions.

Table 1 shows that the optimal signal-to-noise is obtained for $\theta_c \sim 12°$ or $p_c \sim 0.3$ just as with the actual BATSE data sets. It also indicates that the expected signal to noise ratio is significantly smaller for $\theta_c = 4°$. Note that the the correlation function signal seen by NP in the BATSE 1B data set is seen to be much stronger at $\theta_c = 4°$ than at larger angles contrary to the expectation for physically matched pairs unless the matched bursts have unusually small position errors.

In order to understand the difference between the correlation function and the pair matching statistic, we need consider the possibilities that the physically matched pairs have position uncertainties larger than or smaller than the average uncertainties. If the physically matched pairs have larger than average uncertainties, then the pair matching statistic has an advantage over the correlation function because burst pairs with accurate positions that are almost close enough to be considered matched will not be included in the noise for the pair matching statistic. Such pairs will contribute to the noise for the correlation function statistic, however. On the other hand, if the physically matched pairs have smaller than average position errors, then the correlation function can have a higher signal to noise than the pair matching statistic if one one picks the optimal $\theta_c$ value. There is no *a priori* way to know which value of $\theta_c$ is optimal, though. One can also calculate the pair matching statistic for subsamples of bursts with the most accurate positions. Thus, pair matching statistic can generally be considered to be superior to the correlation function for testing models of repeating bursts especially if the repeating bursts are among the fainter bursts in the sample. We should also point out that the $Q_1(0)$ statistic is somewhat more sensitive than the $Q_0(0.3)$ statistic that we have used for this comparison.

The comparison with the nearest neighbor statistic is more complicated because the nearest neighbor statistic does not consider burst pairs in the same way that the pair matching statistic and correlation function statistic do. Because of this, we must specify more details about our artificial burst model. We will consider a variety of models which have the same number of matched burst pairs but different numbers of bursts sources and bursts per source. For example, 60 burst pairs could be generated by 60 sources with 2 bursts per source, 20 sources with 3 bursts per source, 10 sources with 4 bursts per source, 6 sources with 5 bursts per source, or 4 sources with 6 bursts per source. QL have suggested that the bursts responsible for the signal they see may be due to multiple repeaters, and Wang and Lingenfelter (1993) have presented evidence that 5 particular bursts in the BATSE 1B data set may be from the same source. Thus, multiple repeater models seem to be favored by those who argue that the BATSE data implies repeating bursts.

In order to test the sensitivity of the nearest neighbor statistic, we have run sets of Monte Carlo simulations with the burst repetition patterns mentioned above assuming position errors of 8.06° as in the correlation function calculations. 10,000 simulated data sets of 260 and 485 burst positions we generated for the null hypothesis of uncorrelated burst positions, and 1000 simulations were run for each of the models which included multiple bursts from the same position. Following QL, we compare the nearest neighbor distribution with the theoretical prediction (Scott and Tout, 1989) using the Kolmogorov-Smirnov (KS) statistic (Press, *et. al.*, 1992). As QL point out, the standard KS significance estimate is not valid for the nearest neighbor statistic, so we determine the significance of the KS $D$ values by comparing to the 10,000 "null hypothesis" simulations. The results of these simulations are summarized in Table 2.

| # of bursts | # of sources | bursts per source | matched pairs | $D(50\%)$ | $P_{\text{RANDOM}}$ |
|---|---|---|---|---|---|
| 260 | 4 | 6 | 60 | 0.0775 | 0.245 |
| 260 | 6 | 5 | 60 | 0.0823 | 0.188 |
| 260 | 10 | 4 | 60 | 0.0922 | 0.096 |
| 260 | 20 | 3 | 60 | 0.1092 | 0.032 |
| 260 | 60 | 2 | 60 | 0.1393 | 0.0022 |
| 485 | 8 | 6 | 120 | 0.0596 | 0.195 |
| 485 | 12 | 5 | 120 | 0.0634 | 0.140 |
| 485 | 20 | 4 | 120 | 0.0697 | 0.081 |
| 485 | 40 | 3 | 120 | 0.0790 | 0.033 |
| 485 | 120 | 2 | 120 | 0.0931 | 0.0071 |

*Table 2.* The median KS $D$ statistic for 1000 Monte Carlo simulations of repeating

burst models is shown. $P_{\rm RANDOM}$) is the fraction of 10,000 Monte Carlo simulated burst datasets with KS $D$ values exceeding the median value.

It is clear from Tables 1 and 2 that the nearest neighbor statistic is less sensitive than $Q_0(p=0.3)$ or the correlation function (evaluated at the optimal $\theta_c$) for all the multiple repeater models. Only in the case of 2 bursts per source is the nearest neighbor statistic more sensitive than $Q_0(p=0.3)$. It is also important to note that the sensitivity of the nearest neighbor statistic decreases for the larger sample while the sensitivity of $Q_0$ and the correlation function improve with larger samples. Therefore, we conclude that the pair matching statistics, $Q_0(0.3)$ and $Q_1(0)$, are to be preferred on the basis of sensitivity. The improvement in sensitivity over the nearest neighbor statistic is largest in models in which several bursts are detected from the repeating sources and as the number of bursts in the sample increases to $\gtrsim 500$ bursts.

## 4. Pair Matching Analysis of the BATSE Datasets

Before we consider the results of applying our statistical measure to the data, let us discuss several different subsets of the data that we will use below. Since our statistic makes use of the position error estimates, we cannot use some of the BATSE 2B bursts which do not have estimated position errors. Rejecting these bursts cuts the BATSE 2B sample from 585 to 485 bursts. None of the 260 BATSE 1B bursts are removed by this cut. QL and NP both considered sub-samples of the catalog defined by the requirement that $\sigma_{\rm stat}$ be less than some value (9° for QL and 4° for NP). Since the nearest neighbor and correlation function statistics that they used ignored the position error bars, this was a way to test to see if the bursts with smaller error bars had correlations at smaller angles as one might expect if the catalog contains repeaters. Since we do make use of the error bars in our analysis, this is not such an important test for our statistics, but we include it here for comparison with QL and NP. 202 and 133 bursts pass the $\sigma_{\rm stat} < 9°$ and the $\sigma_{\rm stat} < 4°$ cuts respectively in the BATSE 1B data set while 397 and 255 bursts pass these cuts in the BATSE 2B catalog.

One potentially important effect that has not been included in many of the gamma ray burst repetition studies is that the sky exposure seen by BATSE is not uniform. This is important because our statistic is sensitive to numbers of physically matched pairs of bursts that can be much smaller than the number of pair matches that occur by chance. Thus, if the non-uniform sky exposure is not properly modeled, the number of "accidental matches" can be underestimated leading to the "detection" of a spurious signal. Because BATSE's sensitivity is not uniform, faint gamma ray bursts are preferentially detected in certain directions, and since the orientation of BATSE is determined by the pointing of the other CGRO instruments at astronomical objects, we might expect that BATSE's orientation is far from random. (For example, the other instruments might spend a large fraction of the time pointing at specific objects such as the Crab pulsar, Cygnus X-1, and the Galactic Center.) Maoz (1994) has looked at this effect in some detail and determined that this exposure effect may be an important source of systematic errors.

An exposure table that can be used to account for BATSE's non-uniform sky exposure has been provided along with the publicly available BATSE data. However, this table is only accurate for those bursts which would be bright enough to trigger the BATSE detectors from any incident angle. The exposure table for fainter bursts would be a function of their brightness and has not been provided with the BATSE data. Therefore, if we want to fully correct for the non-uniform sky exposure, we must remove the faintest bursts from our sample. We accomplish this by making use of use the threshold table that has been provided with the BATSE 2B catalog. This table gives the trigger efficiency as a function of burst flux for the timescales of each of the BATSE trigger timescales. We have used this table to select flux thresholds for which BATSE has better than 98% detection efficiency. (This means that at least 98% of all bursts not obscured by the Earth with a flux larger than this limit could be seen by BATSE.) Bursts which did not exceed these thresholds for any of the three trigger timescales have not been included in this sample. Our "exposure corrected" sample was corrected for sky exposure using the BATSE sky exposure table, and the "overwrite" bursts were removed from the sample because the sky exposure table does not include the time when these bursts were observed. There are 150 BATSE 1B bursts and 273 BATSE 2B bursts which pass this cut.

Fig. 2 shows the variation of the $Q_1(p)$ statistic with $p$ for our 4 data samples from the BATSE 2B catalog. The $p$ dependence is somewhat more obvious if we plot the differential version of the statistic, i.e., $\Delta Q_1(p) \equiv Q_1(p) - Q_1(p + \Delta p)$ where $\Delta p$ is the bin size. ($\Delta p = 0.1$ in Fig. 2.) Also plotted are the expected signal from a sample which has $10^{-3}$ of all burst pairs originating from the same source and the Monte Carlo average of data sets with random positions. Clearly, none of these data sets shows a significant signal at the $\sim 10^{-3}$ level for all $p$, but some come close to the $10^{-3}$ signal curve for $p \gtrsim 0.7$.

Note that the fraction of all bursts pairs originating from the same source is a somewhat unconventional number to use when describing possible repeating burst signals. For the pair matching statistic (as well as for the correlation function), however, it is the most appropriate since these statistics are burst-pair statistics. For a sample of 485 bursts, $10^{-3}$ corresponds to about 120 bursts pairs which could be generated by as many as 120 sources bursting twice or one source which generates 16 bursts. Thus, $10^{-3}$ burst pairs could correspond to anywhere between 3% and 25% of all 485 bursts being repeats of previous bursts.

Since our main question is whether there is *any* statistically significant signal of burst recurrence in the same position, the fraction of simulated BATSE catalogs with random burst positions that have $Q_1(p) \geq$ the observed value is of particular interest. These values are plotted in Fig. 3 for both the BATSE 1B and 2B catalogs for both "matches" and "antipode matches." Several important features are apparent in Fig. 3. For the "match" statistics for the $\sigma_{\text{stat}} < 9°$ samples (and to a lesser extent in some of the other samples), there appears to be a signal that may be statistically significant for $p \gtrsim 0.6$ which becomes markedly *less* significant at small $p$ where the signal to noise should be larger. From Fig. 2(c), we can see that this is because there is an enhancement of pairs with match probabilities $p_{ij} \gtrsim 0.7$ in both the 1B and 2B $\sigma_{\text{stat}} < 9°$ samples, but for $0.5 < p_{ij} < 0.7$ there is a large deficit of pairs that causes $Q_1(0.5)$ to be about a factor of 20 less significant than $Q_1(0.7)$ for both the 1B and 2B samples. The behavior of $Q_1(p)$ seems consistent with the signal seen by NP's correlation function analysis at $\theta_c < 4°$ and with the nearest neighbor signal seen by QL. However, it is not what we would expect to see if these signals were due to multiple burst repetitions from the same location because matched pairs should be uniformly distributed in $p$. Instead, it appears that there is an excess of pairs separated by angles smaller than the separation that their error bars would indicate.

Because $Q_1(0)$ is predicted (see Fig. 1) to have the best signal to noise, we use it to set our limits on the possibility of gamma ray burst pairs coming from the same sources. In Table 3, we list the probability that a $Q_1(0)$ value at least as large as the observed value is obtained in 10,000 Monte Carlo simulations of the BATSE catalogs assuming random burst positions. These are tabulated along with best fit number of matched pairs according to eq. (2.5). The error bars are the RMS deviation from the mean of the 10,000 simulated BATSE data sets. We note that the best fit number of matched pairs is somewhat sensitive to the assumed Gaussian errors while the match significance values are not. It is obvious from Table 3 and Fig. 3 that the statistical significance of every signal is less for the BATSE 2B catalog than for the BATSE 1B catalog whereas Fig. 1 suggests that a real signal should probably have grown stronger. It is also interesting to note that the "exposure corrected" sample shows no significant signal for "matches" or "antipode matches" in either of the BATSE 1B or 2B data sets. It is tempting to conclude that *all* of the "match" and "antipode match" statistics that are apparently significant in both data sets may be due to the effects of the non-uniform BATSE exposure. The signals are certainly small enough that the number of "accidental matches" is larger than the signal in every case, and the symmetry of the BATSE detector arrangement should be able to explain the existence of an "antipode" signal (see Maoz, 1994). However, it is also true that the exposure corrected sample would be less sensitive possible repetition of the faintest bursts. Thus, it would be worthwhile to repeat this analysis using a model of the BATSE sky coverage that extends to the faintest detectable bursts.

| Sample: | All | $\sigma_{\text{stat}} < 9°$ | $\sigma_{\text{stat}} < 4°$ | Exposure corrected |
|---|---|---|---|---|
| $N_{\text{match}}$ 1B | $13 \pm 25$ | $27 \pm 14$ | $1 \pm 7$ | $3 \pm 12$ |
| $N_{\text{antipode}}$ 1B | $54 \pm 25$ | $24 \pm 14$ | $17 \pm 7$ | $9 \pm 12$ |
| match significance 1B | 29.3 % | 2.65% | 43.36% | 36.96% |
| antipode significance 1B | 2.42% | 4.54% | 1.32% | 21.92% |
| $N_{\text{match}}$ 2B | $-23 \pm 45$ | $0 \pm 26$ | $-13 \pm 13$ | $-9 \pm 20$ |
| $N_{\text{antipode}}$ 2B | $31 \pm 45$ | $39 \pm 26$ | $29 \pm 13$ | $1 \pm 20$ |
| match significance 2B | 68.31% | 48.33% | 82.39% | 65.61% |
| antipode significance 2B | 23.34% | 7.84% | 1.90% | 46.72% |

*Table 3.* The number of matches and anti-matches measured with the $Q_1(0)$ statistic with 1-$\sigma$ error bars determined from the RMS deviation from the mean of 10,000 simulated BATSE catalogs. The significance refers to the fraction of the 10,000 simulated catalogs with larger values of $Q_1(0)$ than the real data.

The only BATSE 2B signal that is apparently significant in Table 3 is the "antipode matching" statistic for the $\sigma_{\text{stat}} < 4°$ sample which appears to be significant at the 98% confidence level. Of course, since we are considering 8 different BATSE 2B data sets, the chance that one should appear significant at the 98% confidence level is $\sim 16\%$. Furthermore, the $\sigma_{\text{stat}} < 9°$ sample has not had all the effects of non-uniform sky exposure removed, so it may be that artifacts related to the non-uniform sky exposure contribute to this signal as discussed above.

In summary, we have developed a powerful new statistical test to search for evidence of repeating gamma ray bursters in the BATSE catalogs. This statistic is shown to be more sensitive (in most cases) than the nearest neighbor and correlation function statistics used previously particularly for faint bursts and multiple burst recurrences. Application of this statistical test to the BATSE 1B and 2B catalogs indicate the signals seen by other authors with the nearest neighbor and correlation function statistics can also been seen with our pair matching statistic. The pair matching statistic reveals that the excess number of burst pairs at small angular distances, seen most strongly in the $\sigma_{\text{stat}} < 9°$ samples, is due to clustering of bursts on scales smaller than would be expected for physical repeating bursts based on the estimated BATSE position errors. When all burst pairs at all positions consistent with repeating bursts are considered, the significance of the pair matching statistic is markedly reduced. There is also no significant matched pair signal in the BATSE 2B data. In addition, we find no significant excess of matched burst pairs or antipode pairs in our exposure corrected sample. This is consistent with the result of Maoz (1994) who has suggested that the signals seen at both small and large angular scales might be do to an exposure effect. We conclude that there is no evidence for repeating gamma ray bursters in the publicly available BATSE data.

## ACKNOWLEDGEMENTS

This work was supported in part by the National Science Foundation through the Center for Particle Astrophysics and by the U.S. Department of Energy at the Lawrence Livermore National Laboratory under contract No. W-7405-Eng-48.

# FIGURE CAPTIONS

1. The expected signal to noise of the pair match statistics $Q_0(p)$ and $Q_1(p)$ for the BATSE 2B catalog. The $Q_1(p)$ curves are the ones which have maximum signal to noise at $p = 0$. The expected signals are given by eq. (2.5), while the noise is the RMS variation from the mean of 10,000 realizations of the BATSE burst catalog in which the burst positions have been randomly reassigned. The normalization of these curves assumes a real signal of $10^{-3}$ of all pairs being matched. The different curves correspond to the different cuts on the data discussed in the text. The signal to noise curves for the BATSE 1B dataset are nearly identical to these except that they are a factor of $\sim 2$ lower.
2. $\Delta Q_1(p) \equiv Q_1(p) - Q_1(p + \Delta p)$ for the BATSE 2B catalog is plotted as a function of $p$ for all bursts, (a), bursts with $\sigma_{\text{stat}} < 9°$, (b), bursts with $\sigma_{\text{stat}} < 4°$, (c), and bursts which pass the "exposure correction" cut, (d). $Q_1(p)$ is obtained from $\Delta Q_1(p')$ by summing over all $p' > p$.
3. The fraction of 10,000 simulated BATSE datasets which have $Q_1(p)$ values which exceed $Q_1(p)$ for the real data is plotted as a function of $p$ for (a) the BATSE 1B catalog, (b) the BATSE 2B catalog, (c) the BATSE 1B data with antipode matching, and (d) the BATSE 2B data with antipode matching.